\documentclass[twocolumn,showpacs,amsmath,amssymb,prb]{revtex4}

\usepackage{epsfig}

\newcommand{\MZO}{Mn$_x$Zn$_{1-x}$O}
\newcommand{\CZO}{Co$_x$Zn$_{1-x}$O}

\begin{document}

\title{Absence of ferromagnetism in Co and Mn substituted polycrystalline ZnO}

\author{G. Lawes}
\email{glawes@lanl.gov}
\affiliation{Los Alamos National Laboratories, Los Alamos, New Mexico, 87545}
\author{A.S. Risbud}
\email{aditi@engineering.ucsb.edu}
\affiliation{Materials Department\\
University of California, Santa Barbara, CA 93106}
 \author{A.P. Ramirez}
\email{apr@lucent.com} \affiliation{Bell Laboratories, Lucent
Technologies, Murray Hill, NJ 07974}
\author{Ram Seshadri}
\email{seshadri@mrl.ucsb.edu}
\affiliation{Materials Department and Materials Research Laboratory\\
University of California, Santa Barbara, CA 93106}

\begin{abstract}
We discuss the properties of semiconducting bulk ZnO when
substituted with the magnetic transition metal ions Mn and Co,
with substituent fraction ranging from $x$ = 0.02 to $x$ = 0.15.
The magnetic properties were measured as a function of magnetic
field and temperature and we find no evidence for magnetic
ordering in these systems down to $T$ = 2 K. The magnetization can
be fit by the sum of a Curie-Weiss term with a Weiss temperature
of $\Theta\gg$100 K and a Curie term. We attribute this behavior
to contributions from both \textit{t}M ions with \textit{t}M
nearest neighbors and from isolated spins.  This particular
functional form for the susceptibility is used to explain why no
ordering is observed in \textit{t}M substituted ZnO samples
despite the large values of the Weiss temperature.
We also discuss in detail the methods we used to minimize any
impurity contributions to the magnetic signal.

\end{abstract}


\maketitle

The ability to control spin as well as electric charge is a
cornerstone of next generation spintronic devices.  Attempts to
inject spin-polarized current into non-magnetic semiconductors
using metallic ferromagnetic contacts have met with mixed
success.\cite{hammar,roukes} Efficient spin injection between
metallic and semiconducting systems appears to be possible only
with an insulating tunnel junction separating the
two.\cite{darryl,rashba,safarov} This complication could be
removed by using a high temperature ferromagnetic semiconductor.
Such a material would play a crucial role in the development of
spintronic devices.  Mn-substituted GaAs has been observed to have
a ferromagnetic transition temperature of up to 172
K,\cite{tanaka} but reports of transition temperatures above 300 K
in Co-substituted TiO$_2$,\cite{matsumoto} among other systems may
be attributed to sample inhomogeneities.\cite{chambers} Recent
work on organic spin-valves\cite{xiong} have shown promising
results, but in this report we concentrate on semiconducting metal
oxides.

One material which shows particular promise for yielding a
suitable ferromagnetic semiconductor is ZnO. Zinc oxide is a wide
bandgap (3.3 eV) semiconductor, so a ferromagnetic version could
also be used as a material for magneto-optical devices. ZnO
substituted with 5\% Mn was predicted to order magnetically above
300 K,\cite{dietl} and the ground state for ZnO substituted with
other transition metal (\textit{t}M) ions is predicted to be
ferromagnetic.\cite{sato} Recently, there has  been work in
investigating the importance of $p$-type doping for producing
ferromagnetic behavior in ZnO.\cite{nicola} There have been a
range of experiments done on Mn-substituted
ZnO,\cite{kolesnik,fukumura,yoon} with some investigation of ZnO
substituted with other transition
metals.\cite{yoon,aditi,brumage,kolesnik3} Several measurements on
Co- and Mn-substituted ZnO grown in thin films have found evidence
for ferromagnetic behavior, but other measurements find no
magnetic transitions. It is thought that clustering of the
magnetic ions into an impurity phase might be responsible for
these features.\cite{kim} One recent report finds evidence for
room temperature ferromagnetism in bulk Mn:ZnO prepared using low
temperature techniques\cite{sharma}. Ferromagnetism has also been
observed in Co:ZnO semiconducting quantum dots\cite{schwartz}, but
the magnetic properties of nanoparticles can differ from bulk
behavior\cite{ganglee}.

In light of these discrepancies in earlier measurements on the
magnetic properties of \textit{t}M substituted ZnO, we have
undertaken a study of bulk \MZO\ and \CZO\ samples which we hope
will resolve some of the ambiguities surrounding reports of
ferromagnetism in these systems.  We investigate bulk samples to
minimize the contribution of extrinsic effects from surface
impurities. Furthermore, by exploring samples with a range of
\textit{t}M fractions prepared using the same technique, we can
investigate systematic trends in these compounds.  Finally, our
measurements are structured to search for and eliminate any
possible contributions from the magnetic impurities which have
plagued earlier studies. Specifically, investigating the
differential high field susceptibility, as described below,
eliminates any potential contributions to the signal from magnetic
impurities not detected by our characterization techniques.  This
is necessary because magnetic measurements are a much more
sensitive probe of ferromagnetic impurities than other
characterization techniques.

In order to investigate the magnetic properties of
ZnO:\textit{t}M, we examined bulk samples of \MZO\, and \CZO\,
with $x$ ranging from 0.02 to 0.15.  It has been found previously
that substitution above these levels leads to the development of a
spinel phase related to Co$_3$O$_4$ in \CZO\,\cite{aditi} so we
confine our experiments to smaller values of $x$. To obtain bulk
\MZO\, and \CZO\, samples, a single-source crystalline precursor
is useful in order to ensure random atomic scale mixing of
Zn$^{2+}$ and Mn$^{2+}$/Co$^{2+}$ ions on lattice sites prior to
decomposition. In addition, decomposition of oxalate precursors
allows the removal of carbon as CO and CO$_2$, leaving a
phase-pure oxide product. Preparation and evidence for the
formation of homogeneous solid solutions in bulk \CZO\/ has been
previously described in detail.\cite{aditi} In a similar manner,
Zn$_{1-x}$Mn$_x$(C$_2$O$_4$)$\cdot$2H$_2$O oxalate precursors were
made with $x$ = 0.02 through $x$ = 0.15. These precursors were
decomposed in air at 1373 K for 15 minutes (with the sample placed
and pulled out of the furnace at temperature).  We performed
elemental X-ray analysis on the Co substituted samples, confirming
the Co fractions.

\begin{figure}
\smallskip
\centering \epsfig{file=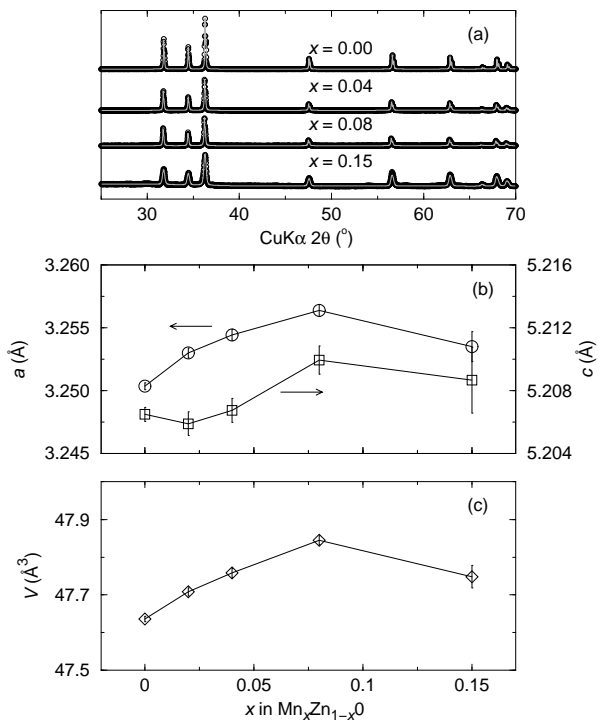, width=8cm} \caption{(a) X-ray
diffraction patterns (experimental data as circles and Rietveld simulations as
lines) of \MZO\, for different $x$ values as indicated in figure.
(b) Evolution of the $a$ and $c$ cell parameters
with increasing $x$ in \MZO.  (c) Evolution of the cell volume with $x$.
In (b) and (c) 3$\sigma$ error bars are indicated.}
\label{fig:xray}
\end{figure}

Powder X-ray diffraction patterns were recorded on a Scintag X2
diffractometer in the Bragg-Brentano configuration using
CuK$\alpha$ radiation.  Data were collected using a step scan of
0.015$^{\circ}$ in 2$\theta$ and subject to Rietveld profile
analysis using the \textsc{xnd} Rietveld code.\cite{berar} Powder
X-ray diffraction data are shown in Fig.~\ref{fig:xray} for the
compositions $x$ = 0.00, 0.04, 0.08 and  0.15 for \MZO. Points are
data and the solid lines are Rietveld fits. None of the samples in
this compositional range showed any evidence for impurity phases.
As the amount of Mn increases, the X-ray peak widths increase
suggesting that crystalline correlation is decreased. This is
consistent with the low natural solubility of Mn$^{2+}$ in the ZnO
lattice. Evolution of lattice parameters, as obtained from the
Rietveld refinement, with $x$ is shown for the \MZO\, in panel (b)
and the evolution of the cell volume in panel (c). The cell
parameter evolution is not simple; the $a$ cell parameter
increases linearly with $x$ only from $x$=0 to $x=0.08$, which
seems to be a limiting composition. The $c$ parameter first
decreases slightly, and then increases. Again, $x$ = 0.08 seems to
be a limiting composition. Substitution of the smaller Zn$^{2+}$
ion (radius = 0.60 \AA) by the larger Mn$^{2+}$ ion (radius = 0.66
\AA) should result in an increase in the unit cell volume. This
increase is once again, systematic till $x$ = 0.08 as seen in
Fig.~\ref{fig:xray}. Slight differences are seen in our results
from those recently published by Kolesnik, Dabrowski, and
Mais,\cite{kolesnik3} who report, for $x$ = 0, 0.05, 0.10, 0.15
and 0.20, that $a$, $c$ and $V$ increase smoothly with $x$ but the
$x$ = 0.15 sample is not single-phase. Because of this, we
restrict our magnetic measurements on \MZO, to samples with $x$
below 0.10.

Magnetic measurements were performed using a commercial
Superconducting Quantum Interference Device (SQUID) magnetometer
(Quantum Design MPMS). For each measurement, we used roughly 30 mg
of sample which gave a signal over three orders of magnitude
larger than the magnetic background of the sample holder.  We
measured both the magnetization as a function of field at fixed
temperature, and the temperature dependence of the susceptibility.
In order to accurately determine the intrinsic magnetization of
the sample as a function of temperature, we measured the
differential susceptibility by subtracting the magnetic moment at
$B$ = 1 T from the moment measured at $B$ =2 T at each
temperature.  The background susceptibility of ZnO was measured
separately and subtracted. This background contribution was
roughly 10\% of the total susceptibility at high temperatures.

Because impurity contributions have previously been misidentified
as intrinsic effects \cite{kim}, it is vital to ensure that we are
unable to detect any impurities in the sample, and also that any
impurity contribution to the magnetization is removed. We know
from XRD that any crystalline impurity fraction must be below the
1\% level. In addition, low field susceptibility measurements as
function of magnetic field at fixed temperature show no deviation
from strictly linear behavior. We take this as evidence for both
the lack of impurities, and that the \textit{t}M ions are fully
substituting into the lattice.  As a final check for impurities,
we measured specific heat of the $x$=0.15 sample (which would be
expected to have the largest impurity contributions), since this
is a bulk probe of magnetic order, even down to very small
impurity fractions.  The absence of any signal in specific heat
arising from long range order in magnetic impurities is consistent
with the magnetic measurements and suggests that we are probing
the intrinsic properties of \textit{t}M-substituted ZnO.
Additionally, by measuring the differential susceptibility as
described above, we can be certain that any spurious contribution
to the signal from ferromagnetic impurity clusters not identified
by the characterizations discussed above will be eliminated.

\begin{figure}
\smallskip
\centering \epsfig{file=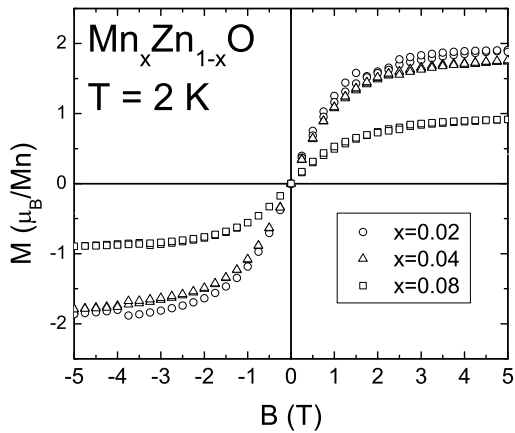, width=8cm} \caption{Magnetic
moment per Mn$^{2+}$ as a function of external field at $T$ = 2 K.
The magnetization of \MZO\, is plotted for four values of $x$
ranging from $x$ = 0.02 to $x$ = 0.08.} \label{fig:MH}
\end{figure}

We plot the magnetization as a function of magnetic field at $T$ =
2 K for \MZO\, for different values of $x$ in figure \ref{fig:MH}.
These magnetization data suggest that there are contributions to
the magnetic signal from both free spins and spins associated with
antiferromagnetic clusters.
  The suppression of the net magnetic moment seen as $x$ is increased is
opposite to expectations based on free Mn impurity spins.  Since
$x$=0.08 is within a factor of two of the 3D percolation threshold
on the ZnO lattice, one would expect to see effects from
clustering. This would lead to a reduction of the measured moment
if the pure Mn end-member were antiferromagnetic.  We observe that
the moment is large for the lowest concentrations ($x$ = 0.02 and
$x$ = 0.04), but becomes smaller as the substituent concentration
increases.   The moment plotted in Figure 2 arises predominantly
from free spins in the system. With increasing $x$, the fraction
of Mn ions belonging to antiferromagnetic clusters which do not
contribute to the magnetic signal increases, which reduces the net
magnetization. This behavior is very similar to that previously
observed in \CZO,\cite{aditi}. Furthermore, the $M(B)$ curves shown
in figure \ref{fig:MH} show no evidence for magnetic hysteresis
even at $T$ = 2 K; there is no ferromagnetic transition in \MZO\,
above this temperature. This observation is in contrast to a
report on room temperature ferromagnetism in \MZO\, at $x$ =
0.02,\cite{sharma} which found evidence for a very small
ferromagnetic moment in samples prepared under relatively low
temperature conditions. The origin for this discrepancy is
unclear, although there are suggestions that these earlier
measurements were sensitive to unreacted manganese
oxides.\cite{kolesnik2}

\begin{figure}
\centering \epsfig{file=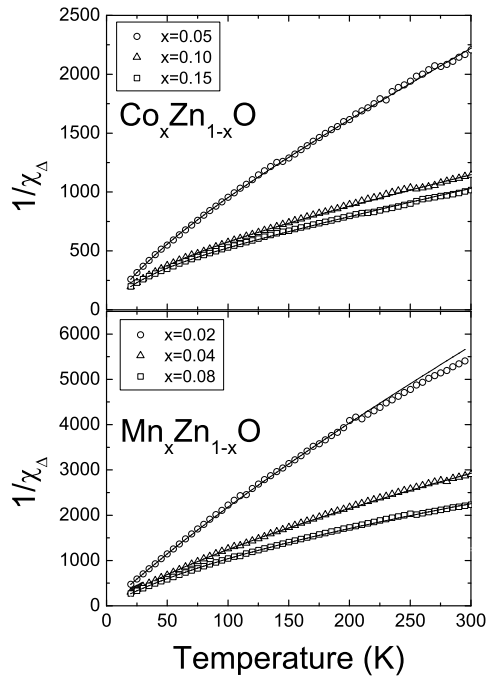, width=8cm} \caption{Inverse
susceptibility for \CZO\/ (upper panel) and \MZO\/ (lower panel)
as a function of temperature. The susceptibility was determined by
looking at the difference in magnetization between $B$ = 2 T and
$B$ = 1 T. The solid lines show fits to the function defined in
Eq. \ref{eq:CWsum}.} \label{fig:CWfig}
\end{figure}

 The temperature dependence of the magnetizations of \MZO\, and \CZO\, are plotted in
 figure \ref{fig:CWfig} as inverse susceptibility versus temperature.
These plots show characteristic behavior observed in other
\textit{t}M-substituted ZnO samples, namely a
 high temperature regime that appears to be close to linear, followed by significant
 curvature at lower temperatures.  Furthermore, there is a systematic variation
 in the high temperature magnetization with substitution fraction.  The samples
 with the smallest values of $x$ show larger values of $1/\chi$, and the
 inverse susceptibility decreases monotonically with increasing $x$ as has
 been observed previously.

 In order to analyze such data, it is typical to make a linear fit of the
 high temperature inverse susceptibility which models Curie-Weiss behavior.
 This approach is motivated by prior work on dilute magnetic semiconductors
  which predicts that the high
 temperature magnetization should follow a modified Curie-Weiss law, where the
 Curie constant and Curie-Weiss temperature are scaled by the substituent
 concentration $x$.\cite{spalek} While this model appears to accurately
 predict the properties of Se and Te based semiconductors even at lower
 temperatures where there are significant deviations from Curie-Weiss
 behavior\cite{spalek,furdyna}, we
 find discrepancies when attempting to apply this formalism to \CZO\, and
 \MZO, particularly in fitting the low temperature susceptibility.
 Furthermore, applying this analysis to ZnO:\textit{t}M samples gives a wide range of very large values
 for the Weiss temperature (from $\Theta_0$ = 960 K to $\Theta_0$ = 1900 K
 for \MZO\cite{kolesnik}) despite the absence of any magnetic order above
 $T$ = 20 K.

To interpret our results, we propose a heuristic model based on
two sets of substituent \textit{t}M spins, motivated by Figure 2
and the subsequent discussion. There is perhaps a hint of such
behavior observed in NMR studies on Cd$_{1-x}$Co$_x$S and
Cd$_{1-x}$Fe$_x$S where it was observed that the splitting of the
$^{113}$Cd spectra can be explained by appealing to different
relative connections to the paramagnetic substituents.\cite{vega}
We assume that one of these subsets of spins (those \textit{t}M
ions with no \textit{t}M nearest neighbors) is completely free, so
the susceptibility follows a simple Curie behavior.  The second
set of spins (those \textit{t}M ions with at least one \textit{t}M
nearest neighbor) is however affected by mean field interactions
with a susceptibility which is expressed as a Curie-Weiss
function. We find that separating the \textit{t}M ions into two
non-interacting subsets, namely free spins and clusters, gives an
excellent fit to the susceptibility over the entire temperature
range.
  We also allow
for the possibility
 that the Curie constants will be different for the two terms.  In
particular, we assume

 \begin{equation}
     \chi=\frac{C_1}{T}+\frac{C_2}{T+\Theta}
     \label{eq:CWsum}
     \end{equation}

 \noindent
 in order to fit the magnetization data for \CZO\, and \MZO.   The solid lines in
 figure \ref{fig:CWfig} show the fits to Eq. \ref{eq:CWsum}, which accurately
 capture both the high temperature behavior and the increasing curvature in
 $\chi_{\Delta}^{-1}$ at lower temperatures.

 Despite these fits requiring three adjustable parameters, this model
 provides insight into the magnetic properties of \CZO\, and \MZO.
 An important observation is that only a subset of the spins are affected by
 magnetic interactions. For Mn-substituted ZnO, the value of
 $\Theta$ lies between roughly +190 K and +360 K for all samples measured.
 This suggests that the spins belonging to \textit{t}M clusters are affected by
antiferromagnetic mean field interactions, consistent with the
$M(B)$ behavior plotted in Figure \ref{fig:MH}. Very similar
 properties are observed for \CZO, but the value of $\Theta$ ranges
 from +160 K ($x$ = 0.05) to +250 K ($x$ = 0.15).

\begin{figure}
\centering \epsfig{file=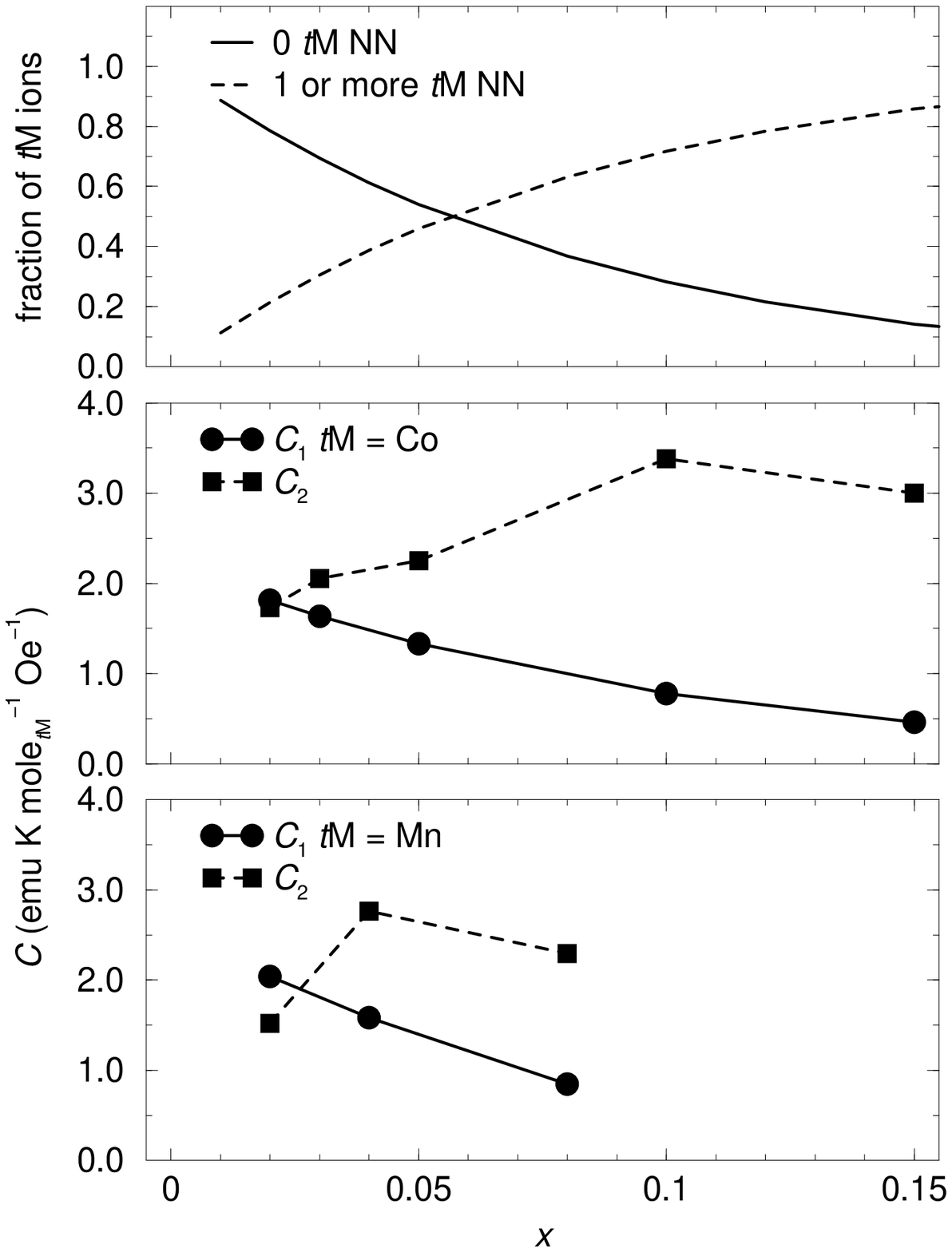, width=8cm} \caption{Values of
the Curie constants for \CZO\/ (middle panel) and \MZO\/ (lower
panel) extracted from the fits to Eq. \ref{eq:CWsum}. The solid
symbols give the Curie constant corresponding to the
 interacting term in Eq. \ref{eq:CWsum}, while the open
symbols correspond to the non-interacting term. Upper Panel:
Computed distribution of $t$M ions on the  cations sites  of the
wurtzite ZnO lattice with  zero $t$M ion near-neighbors (solid
line)  and with one or more $t$M ion near-neighbors (dashed line),
as a function of $x$. We assume random substitution of $t$M on the Zn
site. } \label{fig:NN}
\end{figure}

We plot the Curie constants ($C_1$ and $C_2$) from these fits as a
function of substituent concentration $x$ in figure \ref{fig:NN},
which are plotted scaled to moles Co or Mn. There is a systematic
increase of the contribution from the interacting term with
increasing $x$, while the pure Curie term shows a corresponding
decrease as the substituent concentration is raised. Since the
explicit $x$ dependence of $C_1$ and $C_2$ has been taken into
account by expressing these parameters in moles of Co or Mn, this
implies that the contribution of the interacting Mn or Co ions
increases with the \emph{fraction} of \textit{t}M ions belonging
to magnetic clusters. Conversely, the relative importance of the
non-interacting contribution from Eq. \ref{eq:CWsum} is reduced as
the proportion of isolated spins decreases with increased
substitution.  This observation justifies our initial assumption
that we can divide the magnetic substituents into two distinct
subsets.  However, it should be noted that the Curie constants we
obtain with this analysis are rather different than what one would
expect for free Mn$^{2+}$ or Co$^{2+}$ spins (4.38 and 1.87 emu
K/mole resp.). At present we don't understand the origin of this
discrepancy, although the degree of orbital quenching could be a
relevant parameter. The top panel of the figure shows the fraction
of \textit{t}M ions with no \textit{t}M ion nearest neighbors
(NN), and the fraction with at least one \textit{t}M NN, computed
from a model with the \textit{t}M ions randomly distributed on the
Zn sites.

This analysis of the magnetic properties of \textit{t}M
substituted ZnO reveals several features.  By investigating a
series of compounds with different \textit{t}M fractions, we find
clear evidence for a systematic clustering of spins as the
proportion of magnetic ions is increased.  Additionally, we find
that the dominant spin-spin interactions are antiferromagnetic, in
agreement with other published results\cite{kolesnik}.  Finally,
we offer a simple explanation as to why these strong
antiferromagnetic interactions do not lead to long range spin
ordering.  If we assume that the spins are localized and have no
itinerant character, the lack of magnetic order arises from
geometrical considerations. The 3D site percolation threshold for
an FCC lattice (having the same numbers of nearest neighbors as
the wurtzite lattice) is 19.5\%\cite{stinchcombe}, which is
significantly higher than the maximum \textit{t}M substitution of
10\% for these samples. Since the concentration of magnetic ions
is well below the percolation threshold, we would not expect to
find magnetic order in these systems, at least at temperatures
commensurate with the interaction strengths (few hundred K).

In summary, our measurements on a series of \CZO\, and \MZO\,
samples with $x$ ranging from 0.01 to 0.15 show no evidence for a
ferromagnetic transition in these systems above $T$ = 2 K.  At low
temperatures, the characteristic dependence of the magnetic moment
in \MZO\/ on $B$ suggests that the dominant interactions are
antiferromagnetic, as observed previously in \CZO\, samples
prepared under similar conditions.\cite{aditi} The important
finding is that susceptibility data in both systems are fitted
very well over a large temperature range by the sum of two
Curie-Weiss functions. The fits yield a high temperature
antiferromagnetic interaction ($\Theta\gg$100 K) which we
associate with NN interactions among the magnetic clusters, while
the susceptibility of the spins with no magnetic nearest neighbors
can be fitted very well by a simple Curie function.  We hope to
verify the presence of two distinct spin populations in
\textit{t}M substituted ZnO using techniques such as EPR or
Mossbauer spectroscopy.

Experimentally, we find that the dominant interactions in these
systems are antiferromagnetic consistent with other recent
measurements on these compounds.\cite{kolesnik,aditi} It is
becoming increasingly difficult to envision a scenario whereby
substitution of ZnO with either Mn or Co would produce a
ferromagnetic ground state, at least at temperatures above a few
K. This is consistent with recent work highlighting the importance
of hole substitution in \CZO\, or \MZO\, to obtain
carrier-mediated ferromagnetism.\cite{nicola}

\acknowledgements{A. S. R. is supported by the National Science
Foundation IGERT program under the award DGE-9987618. We
gratefully acknowledge support for the UCSB/LANL collaboration
from the CARE program of the University of California.
The work was also partially supported by the MRL program of
the National Science Foundation under the Award No. DMR00-80034
and by the LDRD program at Los Alamos National Laboratory.
Tsuyoshi Kimura and Nicola Spaldin are thanked for suggestions,
discussions and encouragement.}

\end{document}